\documentclass[showpacs,prd]{revtex4}
\usepackage{epsfig}
\usepackage{graphicx}
\usepackage{dcolumn}
\usepackage{amsmath}
\usepackage{latexsym}

\begin{document}

\title{Path integral action of a particle in a magnetic field in the noncommutative plane and the Aharonov-Bohm effect}  
\author{Sunandan Gangopadhyay$^{a,b,c}$\footnote{e-mail:sunandan.gangopadhyay@gmail.com}, 
Frederik G Scholtz $^{a,d}$\footnote{e-mail:
fgs@sun.ac.za}}
\affiliation{$^a$National Institute for Theoretical Physics (NITheP), 
Stellenbosch 7602, South Africa\\
$^b$Department of Physics, West Bengal State University, 
Barasat, Kolkata 700126, India\\
$^c$Visiting Associate in Inter University Centre for Astronomy $\&$ Astrophysics (IUCAA), Pune 411007, India\\
$^d$Institute of Theoretical Physics, 
Stellenbosch University, Stellenbosch 7602, South Africa}

\begin{abstract}
\noindent The formulation of noncommutative quantum mechanics as a quantum system represented
in the space of Hilbert-Schmidt operators is used to systematically derive, using the standard time slicing procedure, the path integral action for a particle moving in the noncommutative plane and in the presence of
 a magnetic field and an arbitrary potential.   Using this action, the equation of motion
and the ground state energy for the partcle are obtained explicitly. The Aharonov-Bohm phase is derived using a variety of methods and several dualities between this system and other commutative and noncommutative systems are demonstrated.  Finally, the equivalence 
of the path integral formulation with the noncommutative Schr\"{o}dinger equation is also established.

\end{abstract}
\pacs{11.10.Nx} 

\maketitle

\noindent The idea of noncommutative spacetime \cite{snyder} and its possible physical
consequences in quantum mechanics \cite{duval}-\cite{sgthesis}, field theories \cite{witten}-\cite{sghazra} as well as their phenomenological implications in the standard model of particle physics \cite{chams}-\cite{lebed} 
has been an active area of research for quite some time. Futhermore, the framework of noncommutative geometry \cite{connes} provides a useful mathematical setting for the analysis of matrix models in string theory \cite{connes1}.
In a recent paper \cite{fgs}, attention has been paid
to the formal and interpretational aspects of noncommutative quantum mechanics. It has been
discussed that noncommutative quantum mechanics can be viewed as a quantum system represented
in the space of Hilbert-Schmidt operators acting on noncommutative configuration space. Based
on this formalism, the path integral formulation and derivation of the action for a particle
moving in a noncommutative plane was done in \cite{sgfgs} using coherent states. The results obtained were indeed found 
to be in agreement with those found by other means  \cite{muthukumar}, \cite{fgs}. It is worth mentioning that the
noncommutative version of the path integral was also derived in \cite{spal} but it did not reflect the breaking of the time reversal symmetry that one would expect. The same trick of using coherent states to define the propagation kernel was employed in \cite{tan}, however, the final expressions for the propagator and the resulting physics were
quite different from \cite{spal}. In \cite{acat}, a phase-space path integral formulation of noncommutative
quantum mechanics was carried out and its equivalence to the operatorial formulation was shown. 
However, the explicit form of the action derived in \cite{sgfgs} was not obtained here.

In this paper, we proceed to derive the path integral representation and the action for a particle
in a magnetic field moving in a noncommutative plane. From this action, we obtain the equation of motion for
the particle and compute the ground state energy for the particle in a magnetic field in the presence of a harmonic
oscillator potential. We then move on to investigate the Aharonov-Bohm effect (studied earlier by comparing the
solutions of the Schr\"{o}dinger equation in the  noncommutative plane in the presence and absence of the magnetic field
\cite{harms}) in the path integral approach. 
Similar studies using the path integral formulation of quantum mechanics has also been made earlier in \cite{tur}.
However, the results obtained are upto linear order in the noncommutative parameter.
In this context, it is observed that when the noncommutative parameter and magnetic field are related in a specific way, the action for a particle in this magnetic field and noncommutative
plane can be mapped to a particle of zero mass moving in a magnetic field in the commutative plane. 
This relation between  the noncommutative parameter and magnetic field has also been found earlier 
in the context of the Landau level problem in the noncommutative plane \cite{nair}, quantum mechanics on the noncommutative torus \cite{poly} and in the study of exotic Galilean symmetry in the noncommutative plane \cite{duval1}.
Further, the relation between the noncommutative parameter and magnetic field leads to a second class constrained system which upon quantization yields a noncommutative algebra. This is in fact an fairly old and well known result \cite{jackiw,jack}, obtained here for the first time on the level of the action within a path integral setting.  The Aharonov-Bohm phase-difference can then be very easily computed by following the standard arguements.  Obtaining this phase-difference in general
(without any specific relation between the magnetic field and noncommutative parameter) is rather subtle and needs to be done with great care. We also substantiate our results (obtained by the path integral approach) by computing this phase-difference through the transportation of the particle in a closed loop around a magnetic field confined in a solenoid. We also find that the action obtained can be mapped to the action for a particle in a harmonic oscillator potential 
(whose frequency is determined by the magnetic field) moving in the noncommutative plane and also to a commutative
Landau problem.
Finally, we discuss the equivalence of the path integral formulation with the noncommutative Schr\"{o}dinger equation.

To begin our discussion, we present a brief review of the formalism of 
noncommutative quantum mechanics developed recently in \cite{fgs} before constructing the path integral representation
of a particle in a magnetic field on the noncommutative plane. It was suggested in these papers that one can give
precise meaning to the concepts of the classical configuration space and the Hilbert space
of a noncommutative quantum system. The first step is to define 
classical configuration space. In two dimensions, 
the coordinates of noncommutative configuration space satisfy the commutation relation 
\begin{equation}
[\hat{x}, \hat{y}] = i\theta
\label{1}
\end{equation} 
for a constant $\theta$ that we can take, without loss of generality, to be positive. 
The above commutation relation is invariant with respect to $SL(2, R)$ transformation in $(\hat{x}, \hat{y})$-plane, in particular to rotations in this plane.  The annihilation and creation operators are defined by
$\hat b = \frac{1}{\sqrt{2\theta}} (\hat{x}+i\hat{y})$,
$\hat{b}^\dagger =\frac{1}{\sqrt{2\theta}} (\hat{x}-i\hat{y})$
and satisfy the Fock algebra $[\hat{b}, \hat{b}^\dagger ] = 1$. 
The noncommutative configuration space is then 
isomorphic to the boson Fock space
\begin{equation}
\mathcal{H}_c = \textrm{span}\{ |n\rangle= 
\frac{1}{\sqrt{n!}}(\hat{b}^\dagger)^n |0\rangle\}_{n=0}^{n=\infty}
\label{3}
\end{equation}
where the span is taken over the field of complex numbers.

The next step is to introduce the Hilbert space
of the noncommutative quantum system, which is taken to be:
\begin{equation}
\mathcal{H}_q = \left\{ \psi(\hat{x},\hat{y}): 
\psi(\hat{x},\hat{y})\in \mathcal{B}
\left(\mathcal{H}_c\right),\;
{\rm tr_c}(\psi^\dagger(\hat{x},\hat{y})
\psi(\hat{x},\hat{y})) < \infty \right\}.
\label{4}
\end{equation}
Here ${\rm tr_c}$ denotes the trace over noncommutative 
configuration space and $\mathcal{B}\left(\mathcal{H}_c\right)$ 
the set of bounded operators on $\mathcal{H}_c$. 
This space has a natural inner product and norm 
\begin{equation}
\left(\phi(\hat{x}, \hat{y}), \psi(\hat{x},\hat{y})\right) = 
{\rm tr_c}(\phi(\hat{x}, \hat{x})^\dagger\psi(\hat{x}, \hat{y}))
\label{inner}
\end{equation}
and forms a Hilbert space \cite{hol}. To distinguish states in the noncommutative
configuration space from those in the quantum Hilbert space, states in the noncommutative configuration space 
are denoted by $|\cdot\rangle$ and states in the quantum Hilbert space by $\psi(\hat{x},\hat{y})\equiv |\psi)$.  
Assuming commutative momenta, a unitary representation of the noncommutative Heisenberg algebra in terms of operators $\hat{X}$, $\hat{Y}$, $\hat{P}_x$ and $\hat{P}_y$ acting on the states of the quantum Hilbert space 
(\ref{4}) is easily found to be 
\begin{eqnarray}
\hat{X}\psi(\hat{x},\hat{y}) &=& \hat{x}\psi(\hat{x},\hat{y})\quad,\quad
\hat{Y}\psi(\hat{x},\hat{y}) = \hat{y}\psi(\hat{x},\hat{y})\nonumber\\
\hat{P}_x\psi(\hat{x},\hat{y}) &=& \frac{\hbar}{\theta}[\hat{y},\psi(\hat{x},\hat{y})]\quad,\quad
\hat{P}_y\psi(\hat{x},\hat{y}) = -\frac{\hbar}{\theta}[\hat{x},\psi(\hat{x},\hat{y})]~.
\label{action}
\end{eqnarray}
The minimal uncertainty states on noncommutative 
configuration space, which is isomorphic to boson Fock space, 
are well known to be the normalized coherent states \cite{klaud}
\begin{equation}
\label{cs} 
|z\rangle = e^{-z\bar{z}/2}e^{z b^{\dagger}} |0\rangle
\end{equation}
where, $z=\frac{1}{\sqrt{2\theta}}\left(x+iy\right)$ 
is a dimensionless complex number. These states provide an overcomplete 
basis on the noncommutative configuration space. 
Corresponding to these states we can construct a state 
(operator) in quantum Hilbert space as follows
\begin{equation}
|z, \bar{z} )=\frac{1}{\sqrt{\theta}}|z\rangle\langle z|.
\label{csqh}
\end{equation}
These states have the property
\begin{equation}
\hat{B}|z, \bar{z})=z|z, \bar{z})~;~\hat{B}=\frac{1}{\sqrt{2\theta}}(\hat{X}+i\hat{Y}).
\label{p1}
\end{equation}
Writing the trace in terms of coherent states (\ref{cs}) and using 
$|\langle z|w\rangle|^2=e^{-|z-w|^2}$ it is easy to see that 
\begin{equation}
(z, \bar{z}|w, \bar{w})=\frac{1}{\theta}tr_{c}
(|z\rangle\langle z|w\rangle\langle w|)=
\frac{1}{\theta}|\langle z|w\rangle|^2=\frac{1}{\theta}e^{-|z-w|^2}
\label{p2}
\end{equation}
which shows that $|z, \bar{z})$ is indeed a Hilbert-Schmidt operator.  
The `position' representation of a state 
$|\psi)=\psi(\hat{x},\hat{y})$ can now be constructed as
\begin{equation}
(z, \bar{z}|\psi)=\frac{1}{\sqrt\theta}tr_{c}
(|z\rangle\langle z| \psi(\hat{x},\hat{y}))=
\frac{1}{\sqrt\theta}\langle z|\psi(\hat{x},\hat{y})|z\rangle.
\label{posrep}
\end{equation}
We now introduce the momentum eigenstates normalised such that $(p'|p)=\delta(p-p')$
\begin{eqnarray}
|p)&=&\sqrt{\frac{\theta}{2\pi\hbar^{2}}}e^{i\sqrt{\frac{\theta}{2\hbar^2}}
(\bar{p}b+pb^\dagger)}~;~\hat{P}_i |p)=p_i |p)\\
p_x&=&{\mbox Re}\,p~,~p_y={\mbox Im}\,p\nonumber
\label{eg}
\end{eqnarray}
satisfying the completeness relation
\begin{eqnarray}
\int d^{2}p~|p)(p|=1_{Q}~.
\label{eg5}
\end{eqnarray}
We now observe that the wave-function of a ``free particle" on the noncommutative plane is given by \cite{sgfgs}
\begin{eqnarray}
(z, \bar{z}|p)=\frac{1}{\sqrt{2\pi\hbar^{2}}}
e^{-\frac{\theta}{4\hbar^{2}}\bar{p}p}
e^{i\sqrt{\frac{\theta}{2\hbar^{2}}}(p\bar{z}+\bar{p}z)}~.
\label{eg3}
\end{eqnarray}
The completeness relations for the position eigenstates $|z,\bar{z})$ (which is an important
ingredient in the construction of the path integral representation) reads
\begin{eqnarray}
\int \frac{dzd\bar{z}}{\pi}~|z, \bar{z})\star(z, \bar{z}|=1_{Q}
\label{eg6}
\end{eqnarray}
where the star product between two functions 
$f(z, \bar{z})$ and $g(z, \bar{z})$ is defined as
\begin{eqnarray}
f(z, \bar{z})\star g(z, \bar{z})=f(z, \bar{z})
e^{\stackrel{\leftarrow}{\partial_{\bar{z}}}
\stackrel{\rightarrow}{\partial_z}} g(z, \bar{z})~.
\label{eg7}
\end{eqnarray}
This can be proved by using eq.(\ref{eg3}) and computing
\begin{eqnarray}
\int \frac{dzd\bar{z}}{\pi}
(p'|z, \bar{z})\star(z, \bar{z}|p)=
e^{-\frac{\theta}{4\hbar^{2}}(\bar{p}p+\bar{p}'p')}
e^{\frac{\theta}{2\hbar^{2}}\bar{p}p'}\delta(p-p')=(p'|p)~.
\label{eg8}
\end{eqnarray}
Thus, the position representation of the noncommutative
system maps quite naturally to the Voros plane.
With the above formalism and the completeness relations for the
momentum and the position eigenstates 
(\ref{eg5}, \ref{eg6}) in place, we now proceed to write down the
path integral for the propagation kernel
on the two dimensional noncommutative plane. This reads (upto constant factors)
\begin{eqnarray}
(z_f, t_f|z_0, t_0)&=&\lim_{n\rightarrow\infty}\int
\prod_{j=1}^{n}(dz_{j}d\bar{z}_{j})~(z_f, t_f|z_n, t_n)\star_n
(z_n, t_n|....|z_1, t_1)\star_1(z_1, t_1|z_0, t_0)~.
\label{pint1}
\end{eqnarray}

The Hamiltonian (acting on the quantum Hilbert space) for a particle in a magnetic field in the presence of a potential on the noncommutative plane reads
\begin{eqnarray}
\hat{H}=\frac{(\hat{\vec{P}} -e\hat{\vec{A}})^2}{2m}+:V(\hat{B}^{\dagger},\hat{B}):
\label{hamil}
\end{eqnarray} 
where $V(\hat{X},\hat{Y})$ is the normal ordered potential expressed
in terms of the annihilation and creation operators ($\hat{B}$, $\hat{B}^{\dagger}$). 
In the symmetric gauge \footnote{Note that
any asymmetric gauge linear in $\hat{x}$, $\hat{y}$ is related to a symmetric one by a gauge transformation
represented by rotation in the  $(\hat{x}, \hat{y})$-plane and hence leads to the same Hamiltonian. }
\begin{eqnarray}
\hat{\vec{A}}=\left(-\frac{B}{2}\hat{Y}, \frac{B}{2}\hat{X}\right)
\label{gauge}
\end{eqnarray} 
(note that $B$ refers here to the magnetic field and not the annihilation operator $\hat{B}$.  This slight abuse of notation will not create confusion in what follows) the above Hamiltonian takes the form
\begin{eqnarray}
\hat{H}=\frac{\hat{\vec{P}}^2}{2m}+\frac{e^2 B^2}{8m}(\hat{X}^2 +\hat{Y}^2)-\frac{eB}{2m}(\hat{X}\hat{P}_y -\hat{Y}\hat{P}_x)+:V(\hat{B}^{\dagger},\hat{B}):~.
\label{hamil1}
\end{eqnarray} 
With this Hamiltonian, we now compute the propagator over a small segment in the
above path integral (\ref{pint1}). With the help of eq(s) (\ref{eg5}) and (\ref{eg3}), we have
\begin{eqnarray}
(z_{j+1}, t_{j+1}|z_j, t_j)&=&(z_{j+1}|e^{-\frac{i}{\hbar}\hat{H}\tau}|z_j)\nonumber\\
&=&(z_{j+1}|1-\frac{i}{\hbar}\hat{H}\tau +O(\tau^2)|z_j)\nonumber\\
&=&\int_{-\infty}^{+\infty}d^{2}p_j~e^{-\frac{\theta}{2\hbar^{2}}\bar{p}_j p_{j}}
e^{i\sqrt{\frac{\theta}{2\hbar^{2}}}\left[p_{j}(\bar{z}_{j+1}-\bar{z}_{j})+\bar{p}_{j}(z_{j+1}-z_{j})\right]}\nonumber\\
&&\times e^{-\frac{i}{\hbar}\tau[\frac{\bar{p}_j p_{j}}{2m}+\frac{e^2 B^2 \theta}{8m}(2\bar{z}_{j+1}z_{j}+1)
+\frac{ieB}{2m}\sqrt{\frac{\theta}{2}}(p_j \bar{z}_{j+1}-\bar{p}_j z_j)-\frac{eB\hbar}{2m}+V(\bar{z}_{j+1}, z_{j})]}+O(\tau^2)~.
\label{pint2}
\end{eqnarray} 
Substituting the above expression in eq.(\ref{pint1}) and computing the star products explicitly, we obtain (apart from a constant factor)
\begin{eqnarray}
(z_f, t_f|z_0, t_0)=&&\lim_{n\rightarrow\infty}\int \prod_{j=1}^{n} (dz_{j}d\bar{z}_{j})
\prod_{j=0}^{n}d^{2}p_{j}\nonumber\\
&&\exp\left(\sum_{j=0}^{n}\left[\frac{i}{\hbar}\sqrt{\frac{\theta}{2}}\left[p_{j}\left\{\left(1-\frac{ieB\tau}{2m}\right)\bar{z}_{j+1}-\bar{z}_{j}\right\}+\bar{p}_{j}\left\{z_{j+1}-\left(1-\frac{ieB\tau}{2m}\right)z_{j}\right\}\right] +\alpha p_{j}\bar{p}_{j}
-\frac{i}{\hbar}\tau V(\bar{z}_{j+1},z_{j})\right]\right.\nonumber\\
&&\left.~~~~~~~~~~~~~~~~+\frac{\theta}{2\hbar^{2}}\sum_{j=0}^{n-1}p_{j+1}\bar{p}_{j}\right)
\label{pint3}
\end{eqnarray} 
where $\alpha=-\left(\frac{i\tau}{2m\hbar}+\frac{\theta}{2\hbar^{2}}\right)$.
Making the identification $p_{n+1}=p_{0}$, the integrand of the above integral
can be cast in the following form :\\
$\exp\left(-\vec{\partial}_{z_{f}}\vec{\partial}_{\bar{z}_{0}}\right)$
$$\times\exp\left(\sum_{j=0}^{n}\left[\frac{i}{\hbar}\sqrt{\frac{\theta}{2}}\left[p_{j}\left\{\left(1-\frac{ieB\tau}{2m}\right)\bar{z}_{j+1}-\bar{z}_{j}\right\}+\bar{p}_{j}\left\{z_{j+1}-\left(1-\frac{ieB\tau}{2m}\right)z_{j}\right\}\right]
+\alpha p_{j}\bar{p}_{j}-\frac{i}{\hbar}\tau V(\bar{z}_{j+1},z_{j})+\frac{\theta}{2\hbar^{2}}p_{j+1}\bar{p}_{j}\right]
\right).$$ The purpose of the boundary operator in the above expression is to cancel an additional coupling 
which has been introduced between $p_0$ and $p_n$. The introduction of this coupling makes it easy to perform the
momentum integral since it is of the Gaussian form $\exp(\sum_{i,j}p_{i}A_{i,j}\bar{p}_j)$, where $A$ is a 
$D\times D$ ($D=n+1=T/\tau$, $T=t_{f}-t_{0}$) dimensional matrix  given by
\begin{eqnarray}
A_{lr}=\alpha\delta_{l,r}+\frac{\theta}{2\hbar^{2}}\delta_{l+1,r}~.
\label{matrix}
\end{eqnarray}
A simple inspection shows that the eigenvalues and the normalised eigenvectors of the
matrix $A$ are given by
\begin{eqnarray}
\lambda_{k}&=&\alpha+\frac{\theta}{2\hbar^2}e^{2\pi ik/D}\quad;\quad k\in[0,n]\nonumber\\
u_{k}&=&\frac{1}{\sqrt{D}}(1\quad e^{2\pi ik/D}\quad e^{4\pi ik/D}....)^{T}~.
\label{evalues}
\end{eqnarray} 
Since the real part of the eigenvalues of $A$ are nonpositive, one can carry out the momentum integral, to obtain
\begin{eqnarray}
(z_f, t_f|z_0, t_0)&=&\lim_{n\rightarrow\infty}N\int\prod_{j=1}^{n}(dz_{j}d\bar{z}_{j})
\exp\left(-\vec{\partial}_{z_{f}}\vec{\partial}_{\bar{z}_{0}}\right)\nonumber\\
&&~~~~~~~~~~~~~~~~~~~~~\times\exp\left(\frac{\theta}{2\hbar^{2}}\sum_{l=0}^{n}\sum_{r=0}^{n}
\left\{\left(1-\frac{ieB\tau}{2m}\right)\bar{z}_{l+1}-\bar{z}_{l}\right\}A^{-1}_{lr}\left\{z_{r+1}-\left(1-\frac{ieB\tau}{2m}\right)z_{r}\right\}\right)\nonumber\\
&&~~~~~~~~~~~~~~~~~~~~~\times\exp\left(-\frac{i}{\hbar}\tau \sum_{j=0}^{n}V(\bar{z}_{j+1},z_{j})\right).
\label{pintegral1}
\end{eqnarray} 
The inverse of the matrix $A$ is easily obtained as
$A^{-1}_{lr}=\sum_{k=0}^{n}\lambda_{k}^{-1}e^{2\pi i(l-r)k/D}$ leading to \footnote{There is a sign error in the inverse of the matrix $A$ in \cite{sgfgs}. However, the final conclusions regarding the spectrum of the harmonic oscillator in the noncommutative plane remains unaltered.}
\begin{eqnarray}
(z_f, t_f|z_0, t_0)=&&\lim_{n\rightarrow\infty}N\int\prod_{j=1}^{n}(dz_{j}d\bar{z}_{j})
\exp\left(-\vec{\partial}_{z_{f}}\vec{\partial}_{\bar{z}_{0}}\right)
\exp\left(\frac{\theta\tau}{2\hbar^{2}T}\sum_{l,r,k=0}^{n}
\tau\left[\dot{\bar{z}}(l\tau)-\frac{ieB}{2m}\bar{z}(l\tau)\right]\left[\alpha+\frac{\theta}{2\hbar^{2}}e^{-\tau\partial_{(r\tau)}}\right]^{-1}\right.\nonumber\\
&&\left.~~~~~~~~~~~~~~~~~~~~~~~~~~~~\times[e^{2\pi i(l-r)k\tau/T}]\times\tau\left[\dot{z}(r\tau)+\frac{ieB}{2m}z(r\tau)\right]\right)\times\exp\left(-\frac{i}{\hbar}\tau \sum_{j=0}^{n}V(\bar{z}_{j},z_{j})+O(\tau^{2})\right)
\nonumber\\
=&&\lim_{n\rightarrow\infty}N\int\prod_{j=1}^{n}(dz_{j}d\bar{z}_{j})\exp\left(-\vec{\partial}_{z_{f}}\vec{\partial}_{\bar{z}_{0}}\right)\nonumber\\
&&\times\exp\left(\frac{\theta}{2\hbar^{2}T}\sum_{l,r,k=0}^{n}
\tau\left[\dot{\bar{z}}(l\tau)-\frac{ieB}{2m}\bar{z}(l\tau)\right]\left[-\frac{i}{2m\hbar}-\frac{\theta}{2\hbar^{2}}\partial_{(r\tau)}+O(\tau)\right]^{-1}\right.\nonumber\\
&&\left.~~~~~~~~~~~~~~~~~~~~~~~~~~~~ \times[e^{2\pi i(l-r)k\tau/T}]\tau\left[\dot{z}(r\tau)+\frac{ieB}{2m}z(r\tau)\right]\right)\times\exp\left(-\frac{i}{\hbar}\tau \sum_{j=0}^{n}V(\bar{z}_{j},z_{j})+O(\tau^{2})\right)
\label{pintegral2}
\end{eqnarray} 
where in the first line, we have used the fact that $z_{l}=z(l\tau)$
and $z_{l+1}-z_{l}=\tau\dot{z}(l\tau)+O(\tau^{2})$. Taking the 
limit $\tau\rightarrow 0$ and performing the sum over $k$, we finally
arrive at the path integral representation of the propagator
\begin{eqnarray}
(z_f, t_f|z_0, t_0)&=&N\exp\left(-\vec{\partial}_{z_{f}}\vec{\partial}_{\bar{z}_{0}}\right)\int_{z(t_0)=z_0}^{z(t_f)=z_f }\mathcal{D}z\mathcal{D}\bar{z}
\exp({\frac{i}{\hbar}S})
\label{pintegral3}
\end{eqnarray} 
where $S$ is the action given by
\begin{eqnarray}
S=\int_{t_{0}}^{t_{f}}dt \left[\frac{\theta}{2}\left\{\dot{\bar{z}}(t)-\frac{ieB}{2m}\bar{z}(t)\right\}\left(\frac{1}{2m}+\frac{i\theta}{2\hbar}
\partial_{t}\right)^{-1}
\left\{\dot{z}(t)+\frac{ieB}{2m}z(t)\right\}-\frac{e^2 B^2 \theta}{4m}\bar{z}(t)z(t)- V(\bar{z}(t),z(t))\right]~.
\label{action_ncqm}
\end{eqnarray} 

We now compute the ground state energy for the particle in a magnetic field and in the presence of a harmonic oscillator
potential $V=\frac{1}{2}m\omega^2(\hat{X}^{2}+\hat{Y}^{2})$ from the above path integral representation of the transition amplitude. 
Using the normal ordered form of this potential
in terms of the creation and annihilation operators, that is $:V:=m\omega^2 \theta \hat{B}^{\dagger}\hat{B}$,
the action (\ref{action_ncqm}) reads
\begin{eqnarray}
S=\int_{t_{0}}^{t_{f}}dt~\theta\left[\frac{1}{2}\left\{\dot{\bar{z}}(t)-\frac{ieB}{2m}\bar{z}(t)\right\}\left(\frac{1}{2m}+\frac{i\theta}{2\hbar}
\partial_{t}\right)^{-1}
\left\{\dot{z}(t)+\frac{ieB}{2m}z(t)\right\}-\left(\frac{e^2 B^2 }{4m}+m\omega^2\right)\bar{z}(t)z(t)\right]~.
\label{action_nchar}
\end{eqnarray} 
The equation of motion following from the above action is of the following form
\begin{eqnarray}
\ddot{z}(t)+i\left\{\frac{eB}{m}\left(1+\frac{eB\theta}{4\hbar}\right)+\frac{m\omega^2 \theta}{\hbar}\right\}\dot{z}(t)+\omega^2 z(t)=0~.
\label{harsol}
\end{eqnarray} 
It is easy to see that the above equation reduces to the equation of motion 
for the particle in the harmonic oscillator in the $B\rightarrow0$ limit \cite{sgfgs}.
Making an ansatz of the solution of the above equation in the form $z(t)\sim e^{-i\gamma t}$
leads to the following ground state energy eigenvalues for the particle  
\begin{eqnarray}
\gamma=\frac{1}{2}\left\{\frac{m\omega^{2}\theta}{\hbar}+\frac{eB}{m}\left(1+\frac{eB\theta}{4\hbar}\right)\pm\sqrt{\left(\frac{m\omega^{2}\theta}{\hbar}+\frac{eB}{m}\left(1+\frac{eB\theta}{4\hbar}\right)\right)^2+4\omega^{2}}\right\}~.
\label{energyeig}
\end{eqnarray} 
In the $\omega\rightarrow0$ limit, the above expression yields the two frequencies for the particle in a magnetic field on the noncommutative plane to be
\begin{eqnarray}
\gamma=\frac{eB}{m}\left(1+\frac{eB\theta}{4\hbar}\right), 0~.
\label{enereigval}
\end{eqnarray} 
This ground state energy, computed from the path integral formalism, matches with
those obtained by the canonical approach as we show in an appendix.   In the $B\rightarrow0$ limit, the above expression yields the two frequencies for a particle in a harmonic oscillator potential on the noncommutative plane \cite{sgfgs}.

With the above results in place, we now move on to study the Aharonov-Bohm effect. To
proceed, we first observe that the action (\ref{action_ncqm}) can be recast in the following form
\begin{eqnarray}
S=\int_{t_{0}}^{t_{f}}dt~\left[\theta m\left(1+\frac{eB\theta}{2\hbar}\right)^2 
\dot{\bar z}(t)\left(1+\frac{i\theta m}{\hbar}\partial_{t}\right)^{-1}\dot{z}(t)+ieB\theta\left(1+\frac{eB\theta}{4\hbar}\right)\dot{\bar{z}}(t)z(t)-V(\bar{z}(t),z(t))\right]~.
\label{action_mag}
\end{eqnarray} 
Setting $V=0$, we find that the above action can be mapped to a particle of zero mass moving in the commutative plane and in a magnetic field given by
\begin{eqnarray}
B=-\frac{2\hbar}{e\theta}.
\label{choice}
\end{eqnarray} 
The above choice for the magnetic field has also been observed earlier in the literature in different contexts \cite{nair},\cite{poly},\cite{duval1},\cite{schapos}, e.g. in \cite{nair},\cite{poly} they were found to be critical values where
the density of states for a charged particle in a magnetic field with a harmonic oscillator potential becomes infinite.

\noindent Indeed, with this choice of the magnetic field:
\begin{eqnarray}
S&=&\frac{ieB\theta}{2}\int_{t_{0}}^{t_{f}}dt~\dot{\bar{z}}(t)z(t)\nonumber\\
&=&-\frac{eB}{4}\int_{t_{0}}^{t_{f}}dt~[\dot{x}(t)y(t)-\dot{y}(t)x(t)]\nonumber\\
&=&\frac{e}{2}\int_{\vec{x}_{0}}^{\vec{x}_{f}}\vec{A}.d\vec{x}
\label{action_map}
\end{eqnarray} 
where the second line is true upto boundary terms. It is evident from the first line that this is a constrained
system with the following second class constraints
\begin{eqnarray}
\Omega_{1}&=&p_x +\frac{eB}{4}y\approx0\nonumber\\
\Omega_{2}&=&p_y -\frac{eB}{4}x\approx0~.
\label{constr}
\end{eqnarray} 
Introducing the Dirac bracket and replacing $\{. ,.\}_{DB}\rightarrow\frac{1}{i\hbar}[. ,.]$ yield the following
noncommutative algebra
\begin{eqnarray}
[x_i, x_j]=-i\frac{2\hbar}{eB}\epsilon_{ij}=i\theta\epsilon_{ij}~;~[x_i , p_j]=\frac{i\hbar}{2}\delta_{ij}~;
~[p_i , p_j]=-i\hbar\frac{eB}{8}\epsilon_{ij}=\frac{i\hbar^2}{4\theta}\epsilon_{ij}~;~(i, j=1, 2)
\label{nc_alg}
\end{eqnarray} 
where we have used eq.(\ref{choice}). It is to be noted that this noncommutativity was observed earlier in \cite{jackiw, jack}
by noting that in the limit $m\rightarrow0$,  the $y$-coordinate is effectively constrained to the momentum canonical conjugate to the $x$-coordinate. However, in the path integral approach, the mass zero limit arises naturally.

With the usual Aharonov-Bohm experimental set up, one can now easily read off the Aharonov-Bohm 
phase-difference $\phi$ from the action (\ref{action_map}) following the discussion in \cite{sakurai} to be
\begin{eqnarray}
\phi=\frac{eBA}{2\hbar}=\frac{e\Phi}{2\hbar}.
\label{phasediff}
\end{eqnarray} 
Here $A$ is the area enclosed by the loop around which the particle is transported and the magnetic field is non-vanishing and, correspondingly, $\Phi$ is the total magnetic flux enclosed by this loop.
  
In general it is, however, not this easy to obtain the Aharonov-Bohm phase-difference
from the action (\ref{action_ncqm}) (for V=0) (without making the choice (\ref{choice}) for the magnetic field) due to the presence of the non-local time derivative operator. 
To proceed in the general case, consider the experimental setup shown in figure 1.
\begin{figure}
\includegraphics[width=10cm,height=10cm]{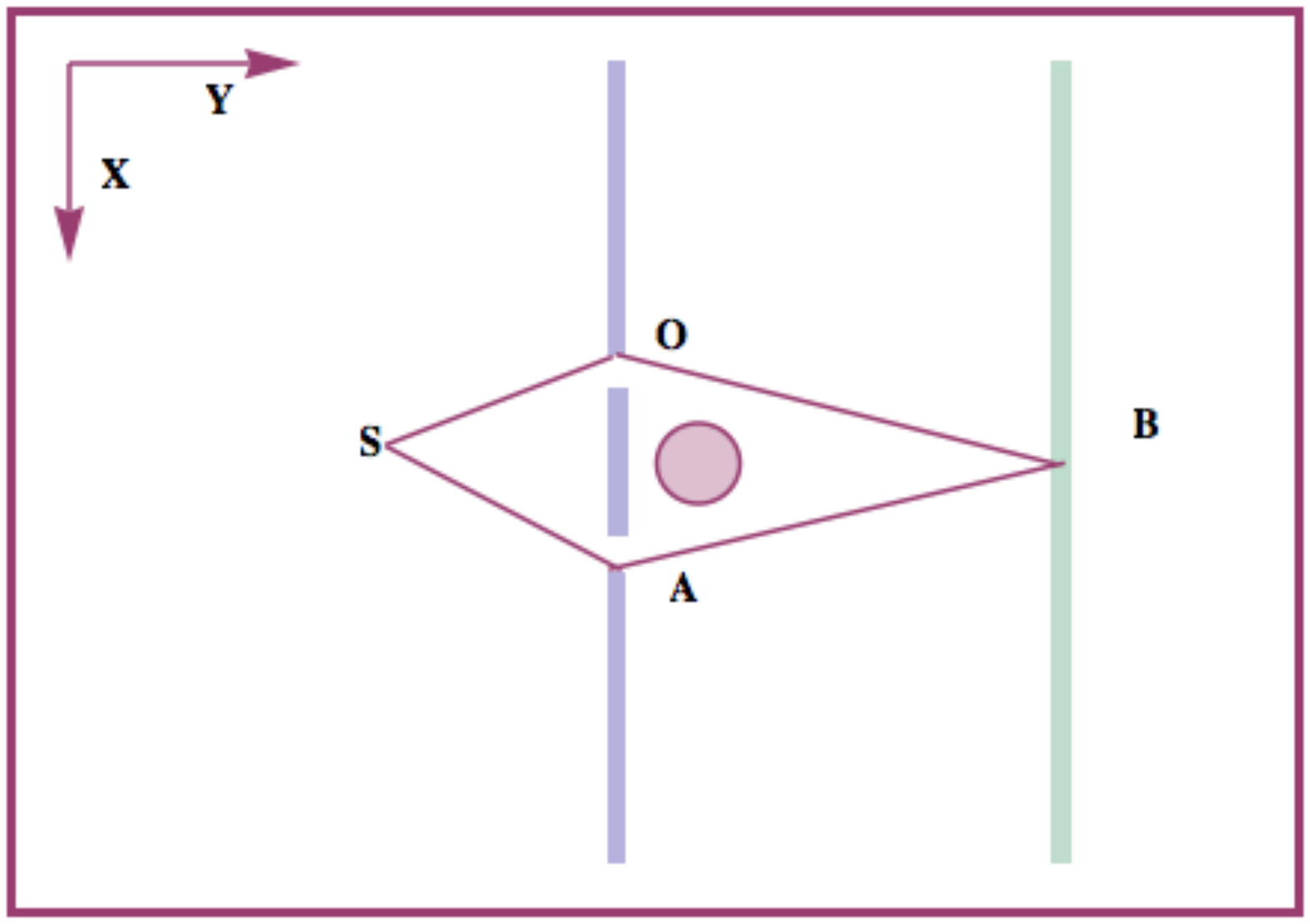}
\caption {Experimental setup for Aharonov-Bohm phase, which shows two slits (A and O) through which particles originating at a source, S, pass.  The two beams pass on opposite sides of a thin solenoid, placed behind the slits and carrying a constant magnetic field, after which they recombine at point B on a screen. }
\end{figure}  
The transition amplitude for a particle originating at S to arrive at point B is the sum over all paths connecting these two points.  From the setup in figure 1, these can naturally be divided into two classes, those paths that pass through slit A and those that pass through slit O and the total transition amplitude therefore consists of the sum of these two contributions.  Since the action is quadratic, the transition amplitude can be computed exactly in a saddle point (or classical) approximation in which case the phase of the transition amplitude is simply given by the action evaluated on the classical path divided by $\hbar$.  Hence, we can compute these two contributions by evaluating them for two classical paths passing through the two slits.  The phase difference is then, of course, obtained by computing the difference between these two phases.  
The result for the action for a classical path starting from the source S ($z_{-T}=\frac{1}{\sqrt{2\theta}}(x_s, -y_s)$ at time $t=-T$), passing through a slit A ($z_2 =\frac{1}{\sqrt{2\theta}}(x_2, 0)$ at time $t=0$)
and reaching a point B ($z_T =\frac{1}{\sqrt{2\theta}}(x_f, y_f)$ at time $t=T$) located on the screen reads
\begin{eqnarray}
S_1 &=& \frac{i\gamma\theta m}{2(1-\cos\gamma T)}\left\{(z_{-T}\bar{z}_2 -c.c) +(\bar{z}_{-T} z_2 e^{i\gamma T}-c.c)
+|z_2|^2 (1-e^{i\gamma T})+|z_{-T}|^2(e^{-i\gamma T}-1)\right.\nonumber\\
&&\left.~~~~~~~~~~~~~~~~~~(\bar{z}_{T}z_2 -c.c) +(z_{T} \bar{z}_2 e^{i\gamma T}-c.c)
+|z_{2}|^2(e^{-i\gamma T}-1) +|z_T|^2 (1-e^{i\gamma T})\right\}
\label{class_action1}
\end{eqnarray} 
where $\gamma=\frac{eB}{m}\left(1+\frac{eB\theta}{4\hbar}\right)$. The result for the action
evaluated on a classical path starting from the same point S (at $t=-T$), passing through a slit 
O ($z_0 =\frac{1}{\sqrt{2\theta}}(0, 0)$ at time $t=0$) and ending at the same point B (at $t=T$)
on the screen can be obtained just by replacing $z_2$ by $z_0$ in the above expression. Taking the difference of these two classical actions and noting that the boundary operator $e^{-\stackrel{\rightarrow}{\partial}_{z_{T}}\stackrel{\rightarrow}{\partial}_{\bar{z}_{-T}}}$ has no effect on this difference, we get
\begin{eqnarray}
S_1 -S_2&=& \frac{m\gamma}{2}\cot\frac{\gamma T}{2}[x_2 (x_2 -x_s -x_f)]+\frac{eB}{\hbar}\left(1+\frac{eB\theta}{4\hbar}\right)\times A.
\label{class_action2}
\end{eqnarray} 
Here $A$ is, as before, the area enclosed by the closed loop and in which the magnetic field is non-vanishing.  Clearly the only topological term in the above expression is the second term, which must be identified with the Aharonov-Bohm-phase.  We remark that in the above calculation one should actually also integrate over the intermediate times at which the particles pass through the slits (to sum over all paths), but one can easily check that this only leads to a multiplicative factor that does not affect the Aharonov-Bohm phase.
The result reduces to eq.(\ref{phasediff}) for the choice of the noncommutative parameter $\theta$ in eq.(\ref{choice}).

An elegant way of obtaining the Aharonov-Bohm-phase is by transporting a particle in a closed loop. 
This can be done by the action of a chain of translation operators on the wave-function as follows
\begin{eqnarray}
e^{-\frac{i}{\hbar}\hat{\pi}_{y}\Delta y}e^{-\frac{i}{\hbar}\hat{\pi}_{x}\Delta x}e^{\frac{i}{\hbar}\hat{\pi}_{y}\Delta y}e^{\frac{i}{\hbar}\hat{\pi}_{x}\Delta x}\Psi~.
\label{loop1}
\end{eqnarray} 
Now using the identity $\hat{S}^{-1}e^{\hat{A}}\hat{S}=e^{\hat{S}^{-1}\hat{A}\hat{S}}$ and
the Baker-Campbell-Hausdorff formula \cite{sakurai}, the above expression can be simplified to
\begin{eqnarray}
e^{\frac{i}{\hbar}\Delta x \Delta y eB\left(1+\frac{eB\theta}{4\hbar}\right)}\Psi~.
\label{loop2}
\end{eqnarray} 
The AB-phase can immediately be read off from the above expression 
and agrees with that obtained from eq.(\ref{class_action2}).

There exists another interesting connection between the action of (\ref{action_ncqm}) and the action of a harmonic oscillator in the noncommutative plane. Setting $V=0$ and making the
following change of variables
\begin{eqnarray}
z(t)=\zeta(t)e^{-\frac{ieB}{2m}t}
\label{harm1}
\end{eqnarray}
eq.(\ref{action_ncqm}) can be recast in the following form
\begin{eqnarray}
S=\int_{t_0}^{t_f}dt~\left[\frac{\theta}{2}e^{\frac{ieB}{2m}t}\dot{\bar\zeta}(t)\left(\frac{1}{2m}+\frac{i\theta}{2\hbar}
\partial_{t}\right)^{-1}\left(\dot{\zeta}(t)e^{-\frac{ieB}{2m}t}\right)-\frac{e^2 B^2 \theta}{4m}
\bar{\zeta}(t)\zeta(t)\right]~.
\label{harm2}
\end{eqnarray}
Now making a Fourier transform of $\dot{\zeta}(t)$, the above expression simplies to
\begin{eqnarray}
S=\int_{t_0}^{t_f}dt~\left[\frac{\theta}{2}\dot{\bar\zeta}(t)\left(\frac{1}{2m}+\frac{eB\theta}{4m\hbar}+\frac{i\theta}{2\hbar}
\partial_{t}\right)^{-1}\dot{\zeta}(t)-\frac{e^2 B^2 \theta}{4m}
\bar{\zeta}(t)\zeta(t)\right]~.
\label{harm3}
\end{eqnarray}
The above action is that of a noncommutative harmonic oscillator with the following identifications
\begin{eqnarray}
\frac{1}{2M}&=&\frac{1}{2m}+\frac{eB\theta}{4m\hbar},\nonumber\\
M\Omega^2 &=&\frac{e^2 B^2}{4m}~.
\label{harm4}
\end{eqnarray}
The equation of motion following from this action reads
\begin{eqnarray}
\ddot{\zeta}(t)+\frac{ie^2 B^2 \theta}{4M\left(1+\frac{eB\theta}{2\hbar}\right)\hbar}\dot{\zeta}(t)+\frac{e^2 B^2}{4M^2 \left(1+\frac{eB\theta}{2\hbar}\right)}\zeta(t)=0~.
\label{harm5}
\end{eqnarray}
The ground state energy of this harmonic oscillator can be obtained by substituting the ansatz $\zeta(t)=e^{-i\Gamma t}$
in the above equation and solving for $\Gamma$ or by simply using the formula for the ground state energy of a 
harmonic oscillator \cite{sgfgs} which yields
\begin{eqnarray}
\Gamma&=&\frac{1}{2\hbar}\left\{M\Omega^2 \theta \pm\Omega\sqrt{M^2 \Omega^2 \theta^2 +4\hbar^2}\right\}\nonumber\\
&=&\frac{eB}{2m}+\frac{e^2 B^2 \theta}{4m\hbar}~,~-\frac{eB}{2m}
\label{harm6}
\end{eqnarray}
where we have used eq.(\ref{harm4}) to obtain the final result. The ground state energy for the problem of the 
particle in a magnetic field in the noncommutative plane can now be obtained (as implied by eq.(\ref{harm1})) by shifting the above energy by $\frac{eB}{2m}$ which gives the result obtained earlier (\ref{enereigval}).

Another interesting link between the problem of a particle moving in a magnetic field in the noncommutative plane and a particle moving in a magnetic field in the commutative plane can be obtained from the following change of variables
\begin{eqnarray}
u=\left(1+\frac{im\theta}{\hbar}\partial_{t}\right)^{-1} z~.
\label{change}
\end{eqnarray} 
Using this we find that the action (\ref{action_mag}) (for $V=0$) 
can be rewritten in the following form
\begin{eqnarray}
S=\int_{t_{0}}^{t_{f}}dt~\theta m\left[\dot{\bar z}(t)\dot{u}(t)+\frac{ieB}{m}\left(1+\frac{eB\theta}{4\hbar}\right)\dot{\bar{z}}(t)u(t)\right]~.
\label{action_mag1}
\end{eqnarray} 
The above action shows that the problem of a particle in a magnetic field
in the noncommutative plane can be mapped to a problem of a particle in a different magnetic field in the commutative plane.  Although this is true on the level of the actions, one must realize that the map between transition amplitudes is more subtle as the boundary conditions on the path integral are also affected by this change of variables.  Indeed, note that the boundary condition on $u$ depends on all higher order derivatives of $z$ at the boundary.  This simply implies the expected, namely, that the commutative transition amplitude is only uniquely determined once all higher order derivatives of $z$ is specified at the boundary.  This is in line with the analysis carried out in \cite{scholtz}.  This does, however,  indicate an interesting duality between a quantum Hall system and a particle in a magnetic field in the noncommutative plane.
Alternatively, one can start from the problem of a particle in a magnetic field $B$ (kept fixed) and get a one
parameter family of problems of a particle in a magnetic field $B^{\star}$ in the noncommutative plane where $B$ and 
$B^{\star}$ are related by
\begin{eqnarray}
B=B^{\star} \left(1+\frac{eB^{\star} \theta}{4\hbar}\right)~.
\label{one_param}
\end{eqnarray}
Solving for $B^{\star}$ gives
\begin{eqnarray}
B^{\star}(\theta)=-\frac{2\hbar}{e\theta}\left\{1-\sqrt{1+\frac{eB\theta}{\hbar}}\right\}~.
\label{solution}
\end{eqnarray}
Expectedly, this solution reduces to the appropriate $B$ and $\theta$ zero limits, i.e. $B^{\star}\rightarrow0$ for
$B\rightarrow0$ and $B^{\star}\rightarrow B$ for $\theta\rightarrow0$.



Finally, we show the equivalence between the path integral formulation in the noncommutative plane and the noncommutative
Schr\"{o}dinger equation. To proceed, we use the fact that the transition amplitude is the propagator which gives the
propagation of the wave-function in the following way
\begin{eqnarray}
\psi(z_f , \epsilon)&=&(z_f , \epsilon|\psi)=\int \frac{d^2 z_i}{\pi}(z_f , \epsilon|z_i , 0)\star_{z_i}(z_i , 0|\psi)\nonumber\\
&=&\int \frac{d^2 z_i}{\pi}(z_f , \epsilon|z_i , 0)\star_{z_i}\psi(z_i, \bar{z}_i , 0)~.
\label{sc1}
\end{eqnarray}
For infinitesimal $\epsilon$ and $\hat{H}=\frac{\hat{P}_{i}^2}{2m}+:V(B^{\dagger}, B):$, the infintesimal
transition amplitude $(z_f , \epsilon|z_i , 0)$ upto $\mathcal{O}(\epsilon)$ reads \cite{sgfgs}
\begin{eqnarray}
(z_f , \epsilon|z_i , 0)=\frac{\theta}{2\hbar^2 \alpha}\left\{1-\frac{i\epsilon}{\hbar}V(\bar{z}_{f}, z_{i})\right\}e^{-\frac{\theta}{2\hbar^2 \alpha}|z_f -z_i|^2}
\label{sc2}
\end{eqnarray}
where $\alpha=\frac{i\epsilon}{2m\hbar}+\frac{\theta}{2\hbar^2}$. Substituting this expression
in eq.(\ref{sc1}), we get
\begin{eqnarray}
\psi(z_f , \epsilon)=\frac{\theta}{2\pi \hbar^2 \alpha}\int d^2 z_i \left\{1-\frac{i\epsilon}{\hbar}V(\bar{z}_{f}, z_{i})\right\}e^{-\frac{\theta}{2\hbar^2 \alpha}|z_f -z_i|^2}\star_{z_i}\psi(z_i, \bar{z}_i , 0)~.
\label{sc3}
\end{eqnarray}
Making a change of variables to
\begin{eqnarray}
z_i = z_f +\eta
\label{sc4}
\end{eqnarray}
the above expression can be recast in the form
\begin{eqnarray}
\psi(z_f , \epsilon)=\frac{\theta}{2\pi \hbar^2 \alpha}\int d^2 \eta \left\{1-\frac{i\epsilon}{\hbar}V(\bar{z}_{f}, z_{f}+\eta)\right\}e^{-\frac{\theta}{2\hbar^2 \alpha}|\eta|^2}\star_{\eta}\psi(z_f +\eta, \bar{z}_f +\bar\eta , 0)~.
\label{sc5}
\end{eqnarray}
Using the form of the star product and the fact that 
$f(z+\eta)=e^{\eta\stackrel{\rightarrow}{\partial_z}}f(z)$, the above equation can be simplified to
\begin{eqnarray}
\psi(z_f , \epsilon)=\frac{\theta}{2\pi \hbar^2 \alpha}\int d^2 \eta~e^{-\frac{\theta}{2\hbar^2 \alpha}|\eta|^2}e^{\frac{i\epsilon}{2m\hbar\alpha}\eta\stackrel{\rightarrow}{\partial}_{z_{f}}}
e^{\bar{\eta}\stackrel{\rightarrow}{\partial}_{\bar{z}_f}}\left[1-\frac{i\epsilon}{\hbar}V(\bar{z}_f , z_f)
e^{\eta\stackrel{\leftarrow}{\partial}_{z_f}} \right]\psi(z_f , \bar{z}_f, 0)~.
\label{sc6}
\end{eqnarray}
We now expand the exponential involving $\epsilon$ in the exponent in a power series (keeping terms upto
$\mathcal{O}(\epsilon)$) and perform the $\eta$ integral to get
\begin{eqnarray}
\psi(z_f , \epsilon)=\psi(z_f , 0) +\frac{i\epsilon\hbar}{m\theta}\frac{\partial^{2}}{\partial_{\bar{z}_f} \partial_{z_f}}\psi(z_f , 0)-\frac{i\epsilon}{\hbar}\frac{\theta}{2\hbar^2 \alpha}V(\bar{z}_f , z_f)\star_{z_f}
\psi(z_f , 0)~.
\label{sc7}
\end{eqnarray}
In the limit $\epsilon\rightarrow0$, we get the time dependent Schr\"{o}dinger equation in NC plane
\begin{eqnarray}
i\hbar\partial_{t}\psi(z_f , t)=\left[-\frac{\hbar^2}{m\theta}\frac{\partial^{2}}{\partial_{\bar{z}_f} \partial_{z_f}}+V(\bar{z}_f , z_f)\star_{z_f}\right]\psi(z_f , t)~.
\label{sc8}
\end{eqnarray}

To summarise: In this paper, we have systematically derived the path integral
representation of the propagation kernel for a particle in a magnetic field in the presence of an arbitrary potential moving in the the noncommutative plane using the recently proposed formulation of noncommutative
quantum mechanics. From the path integral, we have obtained the action
for the particle in noncommutative quantum mechanics. This is one of the 
important results in our paper. The equation of motion of the particle is obtained from this action and was used
to compute the ground state energy of the particle and to show that the result
is in conformity with the results obtained by other methods.
We then investigated the Aharonov-Bohm effect using the path integral formulation. In this context we
found an interesting connection (observed earlier in different contexts in \cite{nair}-\cite{duval1}) with a particle of zero mass moving in a magnetic field related in a particular way to the noncommutative parameter. The  Aharonov-Bohm phase-difference in this case is easy to read off.  Although
the computation is in general a bit more subtle, this computation has also been performed and found to be in agreement with other techniques of computation. We also observed interesting connections of this action with those of a noncommutative harmonic oscillator and a commutative Landau problem. Finally, we discussed the equivalence of the path integral formulation and the noncommutative Schr\"{o}dinger equation. \\



\section*{Appendix}
\noindent In this appendix, we obtain the energy spectrum of the particle in a magnetic field in the noncommutative
plane by the canonical method \cite{mezincescu}.
The noncommuting coordinates can be expressed in terms of commuting coordinates and their momenta in the form
\begin{eqnarray}
\hat{X}&=&X-\frac{\theta}{2\hbar}P_{y}\nonumber\\
\hat{Y}&=&Y+\frac{\theta}{2\hbar}P_{x}~;~\hat{P}_{i}=P_i.
\label{app1}
\end{eqnarray}
Under this change of variables, the Hamiltonian (\ref{hamil1}) can be rewritten as
\begin{eqnarray}
\hat{H}=\frac{1}{2m}\left\{h_{1}^{2}\vec{X}^{2}+h_{2}^{2}\vec{P}^{2}+h_{3}(XP_{y}-YP_{x})\right\}
\label{app2}
\end{eqnarray}
where
\begin{eqnarray}
h_{1}^{2}&=&m^2 \omega_{c}^2 ~;~\omega_{c}=\frac{eB}{2m}\nonumber\\
h_{2}^{2}&=&\left(1+\frac{m\omega_{c}\theta}{2\hbar}\right)^2 \nonumber\\
h_{3}&=&2m\omega_{c}\left(1+\frac{m\omega_{c}\theta}{2\hbar}\right)=2h_1 h_2~.
\label{app3}
\end{eqnarray}
Introducing the operators
\begin{eqnarray}
a_{x}=\frac{ih_2 P_x +h_1 X}{\sqrt{h_3 \hbar}}~;~a_{y}=\frac{ih_2 P_y +h_1 Y}{\sqrt{h_3 \hbar}}
\label{app4}
\end{eqnarray}
where $[a_x , a_{x}^{\dagger}]=1=[a_y , a_{y}^{\dagger}]$
leads to
\begin{eqnarray}
\hat{H}=\frac{h_3 \hbar}{2m}\{(a_{x}^{\dagger}a_x +a_{y}^{\dagger}a_y +1 )+i(a_{y}^{\dagger}a_x -a_{x}^{\dagger}a_y)\}~.
\label{app5}
\end{eqnarray}
Finally, defining
\begin{eqnarray}
a_{\pm}=\frac{a_x \pm i a_y}{\sqrt{2}}
\label{app6}
\end{eqnarray}
where $[a_+ , a_{+}^{\dagger}]=1=[a_{-} , a_{-}^{\dagger}]$
leads to the following second quantized Hamiltonian from which the energy spectrum can easily be read off:
\begin{eqnarray}
:\hat{H}:&=&\frac{h_3 \hbar}{m}a_{-}^{\dagger}a_{-}=\frac{eB}{m}\left(1+\frac{eB\theta}{4\hbar}\right)\hbar a_{-}^{\dagger}a_{-}~.
\label{app7}
\end{eqnarray}
For the sake of completeness, we also obtain the Heisenberg equations of motion for the noncommutative
variable $\hat{Z}$ which reads 
\begin{eqnarray}
\dot{\hat Z}(t)&=&\frac{1}{i\hbar}[\hat Z, \hat H]\nonumber\\
&=&\frac{1}{m}\left(1+\frac{eB\theta}{2\hbar}\right)\left\{\frac{\hat{P}}{\sqrt{2\theta}}-\frac{ieB}{2}\hat Z\right\}~.
\label{he1}
\end{eqnarray}
Similarly, one gets
\begin{eqnarray}
\dot{\hat P}(t)&=&-\sqrt{2\theta}\frac{e^2 B^2}{4m}\hat{Z}-\frac{ieB}{2m}\hat{P}~.
\label{he2}
\end{eqnarray}
Differentiating eq.(\ref{he1}) and combining it with eq(s)(\ref{he1}, \ref{he2}) leads to
\begin{eqnarray}
\ddot{\hat Z}(t)+\frac{ieB}{m}\left(1+\frac{eB\theta}{4\hbar}\right)\dot{\hat{Z}}=0
\label{he3}
\end{eqnarray}
which is the equation of motion for the particle.
\vskip 0.5 cm
\noindent {\bf{Acknowledgements}} : This work was supported under a grant of the National Research Foundation of South Africa. The authors would also like to thank the referees for useful comments.


\end{document}